\documentclass[aps, prb, amsmath, amssymb, twocolumn, a4paper, superscriptaddress]{revtex4-2}
\usepackage{graphicx} 
\usepackage{dcolumn} 
\usepackage{bm} 
\usepackage{hyperref} 
\usepackage{color} 
\usepackage{gensymb} 
\makeatletter
\g@addto@macro\bfseries{\boldmath}
\makeatother
\def\musr{$\mu^+$SR}
\def\neel{N{\'e}el}

\begin{document}
\title{Depth-dependent magnetic crossover in a room-temperature skyrmion-hosting multilayer}
\author{T.~J.~Hicken}
\affiliation{Department of Physics,  Centre for Materials Physics, Durham University, Durham, DH1 3LE, United Kingdom}
\affiliation{Department of Physics, Royal Holloway, University of London, Egham, TW20 0EX, United Kingdom}
\affiliation{Paul Scherrer Institute, Forschungsstrasse 111, 5232 Villigen PSI, Switzerland}
\author{M.~N.~Wilson}
\affiliation{Department of Physics,  Centre for Materials Physics, Durham University, Durham, DH1 3LE, United Kingdom}
\affiliation{Department of Physics and Physical Oceanography, Memorial University of Newfoundland, St. John's, NL A1B 3X7, Canada}
\author{Z.~Salman}
\affiliation{Paul Scherrer Institute, Forschungsstrasse 111, 5232 Villigen PSI, Switzerland}
\author{S.~L.~Zhang}
\affiliation{School of Physical Science and Technology, ShanghaiTech University, Shanghai, 201210, China; ShanghaiTech Laboratory for Topological Physics, ShanghaiTech University, Shanghai 200031, China}
\author{S.~J.~R.~Holt}
\affiliation{Max Planck Institute for the Structure and Dynamics of Matter, Luruper Chaussee 149, 22761 Hamburg, Germany}
\author{T.~Prokscha}
\author{A.~Suter}
\affiliation{Paul Scherrer Institute, Forschungsstrasse 111, 5232 Villigen PSI, Switzerland}
\author{F.~L.~Pratt}
\affiliation{ISIS Pulsed Neutron and Muon Facility, STFC Rutherford Appleton Laboratory, Harwell Oxford, Didcot, OX11 OQX, United Kingdom}
\author{G.~van~der~Laan}
\affiliation{Diamond Light Source, Harwell Science and Innovation Campus, Didcot, Oxfordshire, OX11~0DE, United Kingdom}
\author{T.~Hesjedal}
\affiliation{Department of Physics, Clarendon Laboratory, University of Oxford, Oxford, OX1 3PU, United Kingdom}
\author{T.~Lancaster}
\affiliation{Department of Physics,  Centre for Materials Physics, Durham University, Durham, DH1 3LE, United Kingdom}

\begin{abstract}
	Skyrmion-hosting multilayer stacks are promising avenues for applications, although little is known about the depth dependence of the magnetism.
	We address this by reporting the results of circular dichroic resonant elastic x-ray scattering (CD-REXS), micromagnetic simulations, and low-energy muon-spin rotation (LE-\musr) measurements on a stack comprising $[$Ta/CoFeB/MgO$]_{16}$/Ta on a Si substrate.
	Energy-dependent CD-REXS shows a continuous, monotonic evolution of the domain-wall helicity angle with incident energy, consistent with a three-dimensional hybrid domain-wall-like structure that changes from \neel-like near the surface to Bloch-like deeper within the sample.
	LE-\musr\ reveals that the magnetic field distribution in the trilayers near the surface of the stack is distinct from that in trilayers deeper within the sample.
	Our micromagnetic simulations support a quantitative analysis of the \musr\ results.
	By increasing the applied magnetic field, we find a reduction in the volume occupied by domain walls at all depths, consistent with a crossover into a region dominated by skyrmions above approximately 180~mT.
\end{abstract}

\date{\today}
\maketitle

\section{Introduction}
Several material classes have emerged that host magnetic skyrmion excitations~\cite{lancaster2019skyrmions,nagaosa2013topological}, but for applications it is necessary to stabilise skyrmions at room temperature.
For this purpose, multilayer systems show promise~\cite{jiang2017skyrmions}.
These comprise a repeated stack of a magnetic layer and one or more heavy elements, e.g. Ta/CoFeB/MgO.
Here, a Dzyaloshinskii-Moriya interaction (DMI) arises from inversion symmetry-breaking at interfaces between the layers, leading to  skyrmions that can exist over a wide range of temperature, extending up to 500~K~\cite{zhang2018creation}.
While skyrmions are often assumed to be a quasi-two-dimensional texture that identically repeats in the third dimension, recent work on Ta/CoFeB/MgO multilayers suggests a  variation of the spin texture across multiple repeats of the multilayer~\cite{li2019anatomy}, with micromagnetic simulations predicting that the helicity changes from \neel-type at the surface, to Bloch-type deeper in the stack.

There are many different techniques with which one can characterise three-dimensional spin textures such as skyrmions, which all have different strengths and weaknesses~\cite{burgos2023probing,flewett2021general,milde2013unwinding,henderson2023three,wolf2022unveiling,seki2022direct,donnelly2021experimental,witte20202d,donnelly2017three,Fert_hybrid_[Ir/Co/Pt]n_REXS_Sciadv_18,grelier2022three}.
In the case of Ta/CoFeB/MgO multilayer stacks, circular dichroic resonant elastic x-ray scattering (CD-REXS) measurements~\cite{sm,Gerrit_REXS_Physique_08,Winding_Natcommun_17} have previous been employed to study the system~\cite{CDREXS2017,DER_skyrmion_18,Depth_profile_18,li2019anatomy}.
In this system, alternating non-magnetic spacers establish a pronounced dipole-dipole interaction among the CoFeB layers, leading to a unique hybrid three-dimensional (3D) domain wall configuration with a Bloch layer that is located in the middle of the stack~\cite{Fert_hybrid_[Ir/Co/Pt]n_REXS_Sciadv_18,li2019anatomy}.
The helicity angle $\chi$, which describes the character of domain walls as well as skyrmions, continuously rotates between the two extremes throughout the  stack, with a characteristic  profile $\chi(z)$, where $z$ is the depth into the sample. 
In this case, the top CoFeB layer favors a \neel-type domain wall with a helicity angle of $\chi = 180\degree$ or $0\degree$, while the middle trilayer that hosts the Bloch point exhibits a Bloch-type domain wall with $\chi=\pm 90^\circ$.
The bottom trilayer is again of \neel-type, but with $\chi$ changed by $180\degree$ from the surface~\cite{li2019anatomy,CoTb_PRL_21}. 
Introducing interfacial DMI  moves the Bloch configuration in $z$~\cite{Fert_hybrid_[Ir/Co/Pt]n_REXS_Sciadv_18,li2019anatomy}.
In nonzero external magnetic fields, a similar 3D magnetization characteristic is also expected for skyrmions.

Despite the predictions from micromagnetic simulations of Ta/CoFeB/MgO multilayer stacks being consistent with the CD-REXS measurements, a direct study of the depth-resolved magnetic environment in the stack has hitherto been lacking.
Here we provide these measurements through an investigation of depth-dependent magnetism in the multilayer $[$Ta(2)/CoFeB(1.5)/MgO(2)$]_{16}$/Ta(5) (thicknesses in nm) using low-energy muon-spin rotation (LE-\musr)~\cite{blundell2021muon,prokscha2008new,morenzoni2000low}.
Previous work~\cite{lancaster2015transverse,franke2018magnetic,hicken2020magnetism,hicken2021megahertz} has shown that bulk \musr~\cite{blundell1999spin,blundell2021muon} is a useful tool in the study of skyrmion systems and their dynamics, while LE-\musr\ has probed depth-dependent magnetism at interfaces~\cite{krieger2019topology} and the behavior of superconducting~\cite{suter2011} and magnetic~\cite{boris2011} superlattices.
However, these studies do not address the evolution of the properties with depth within a repeating structure.
Combining our  measurements on $[$Ta(2)/CoFeB(1.5)/MgO(2)$]_{16}$ with energy-resolved CD-REXS, and interpreting them with support from micromagnetic simulations, we show here how the domain-wall angle and magnetic structure varies with depth, and reveal a crossover in magnetic properties as a function of depth.

\section{Experimental and Computational Details}
The $[$Ta(2)/CoFeB(1.5)/MgO(2)$]_{16}$/Ta(5) (thicknesses in nm) multilayered thin films were grown by magnetron sputtering onto Si wafers~\cite{li2019anatomy, Burn2020}.
A Ta layer was always chosen as the final layer to prevent oxidation.
The base pressure of the deposition system was $<$3$ \times 10^{-6}$~Pa.
The sputter gas, Ar, was used at a pressure of 0.3~Pa.
The deposition rates for Ta, CoFeB, and MgO were 0.48, 0.30, and 0.05 \AA/s respectively.
A sputtering power of 100~W was used.

The CD-REXS experiments were carried out in the RASOR diffractometer on beamline I10 at the Diamond Light Source (UK).
The CD-REXS patterns shown are obtained from the difference between the scattering intensities for left- and right-circularly polarized x-rays for varying photon energies around the Fe $L_3$ absorption edge.
The dichroism extinction condition directly reveals the twisting angle of the magnetic structure averaged over the probed depth.

Micromagnetic simulations were performed using the Ubermag package~\cite{fangohr2024vision}, with \textsc{oommf}~\cite{donahue1999oommf} as the computational backend.
The same material parameters were used as in Ref.~\cite{li2019anatomy}, with the layer thicknesses changed to be appropriate for our sample.
The parameters included are exchange $A=10$~pJm$^{-1}$, interfacial (C$_{nv}$) DMI $D=0.4$~mJm$^{-2}$, and out-of-plane uniaxial anisotropy $K_\text{u}=0.9$~MJm$^{-3}$, and the calculations also included demagnetisation in the Hamiltonian.
In magnetic layers, the magnetic saturation was $M_\text{s}=1.18$~MAm$^{-1}$.
Open boundary conditions were employed on samples of lateral size $300\times300$~nm, and the system used a discretization of $4\times4\times0.5$~nm.
The complete code used can be found in the Supplemental Material~\cite{sm}.

Low-energy (LE)-\musr\ measurements of were carried out at the Swiss Muon Source, Paul Scherrer Institut, Switzerland, using the LEM instrument, which allows the implantation depth of the muons to be altered by changing the energy of the incoming muons~\cite{morenzoni2000low,prokscha2008new}.
Measurements were performed at 290~K, either in the ZF or TF geometry.
Data analysis was carried out using both the WiMDA program~\cite{pratt2000wimda}, making use of the MINUIT algorithm~\cite{james1975minuit} via the iminuit~\cite{iminuit} Python interface for global refinement of parameters, and musrfit~\cite{suter2012musrfit}.
Stopping profiles of the muons in the multilayer stack were modeled using the TRIM.SP software~\cite{morenzoni2002implantation}.
A model of the multilayer stack was made in the software by inserting layers matching the density of each of the materials at the correct thickness to match the experimental system.
TRIM.SP then calculates histograms of the stopping profiles for a particular incident muon energy; we varied this parameter to investigate the changes in stopping profiles.
For these calculations, at all incident energies, the energy variance was set to 0.45~keV, and muons were incident perpendicular to the surface with a variance of 15$\degree$.

\section{Results \& Discussion}
\subsection{Circular dichroic resonant elastic x-ray scattering}
\begin{figure}
	\includegraphics[width=\linewidth]{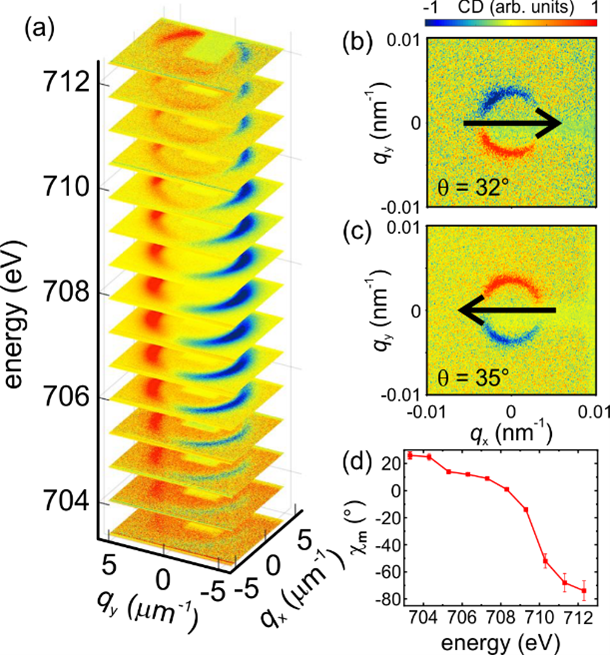}
	\caption{CD-REXS patterns of the multilayered sample, measured at room temperature in zero magnetic field. (a) Energy-dependent CD-REXS patterns measured at $\theta=12^\circ$. The individual patterns at the energies at which a helicity angle could be extracted are shown in Ref.~\cite{sm}. (b,c) CD-REXS patterns measured at $\theta=32^\circ$ and $35^\circ$, respectively, using a photon energy of 707.8\,eV (at the Fe $L_3$ edge). (d) Energy dependence of the measured helicity angle $\chi_\text{m}$ across the $L_3$ absorption edge.}
	\label{fig:CDREXS}
\end{figure}

We studied the depth-dependent character of the magnetic domain walls in the multilayer system $[$Ta(2)/CoFeB(1.5)/MgO(2)$]_{16}$/Ta(5) on a Si substrate using CD-REXS~\cite{CDREXS2017, DER_skyrmion_18, Depth_profile_18, van2014x}, which is sensitive to the twisting angle of spin spirals and skyrmions, and was previously applied to retrieve $\chi$ of a magnetic twisted domain state~\cite{DER_skyrmion_18}. 
Figure~\ref{fig:CDREXS}(a) shows energy-dependent dichroic scattering patterns, obtained on the multilayer sample at room temperature and in zero applied magnetic field. 
The ring-like magnetic scattering patterns were obtained by integrating several azimuthal angles using the $\phi$-axis of the diffractometer.   
The dichroic patterns feature a dividing vector that separates the blue and red regions of negative and positive dichroic contrast, governed by the dichroism extinction rule~\cite{CDREXS2017}, see Fig. S2 in Ref.~\cite{sm}. 
The azimuthal direction of the dividing vector uniquely reveals the value of $\chi$~\cite{DER_skyrmion_18} (Fig. S2 in Ref.~\cite{sm}).
For multilayer systems with a non-uniform $\chi(z)$ profile, the measured $\chi_\text{m}$ is the average $\chi$ from all trilayers with different weightings.  
Experimentally, by varying the photon energy across the Fe $L_3$ edge, or the x-rays incidence angle $\theta$, the weight changes due to the varied sampling depth of the soft x-rays, leading to a different measured average $\chi_\text{m}$~\cite{Depth_profile_18, CoTb_PRL_21, Binding_String_NL_22}.
For the investigated $[$Ta(2)/CoFeB(1.5)/MgO(2)$]_{16}$/Ta(5) heterostructure, the attenuation length of the x-rays, $\Lambda$, is 79~nm at the Fe $L_3$ edge (707.8~eV) at normal incidence, and 152~nm off-edge (700~eV).
The sampling depth varies with the angle between the x-ray beam and surface normal, $\alpha$, as $\Lambda\cos(\alpha)/2$~\cite{Twist2021}.
A systematic variation of these  parameters therefore provides a strategy for investigating the depth-dependent structural property of a 3D hybrid domain wall.

As can be seen in Fig.~\ref{fig:CDREXS}, the direction of the dividing vector undergoes a gradual change across the photon energy spectrum.
Due to the complexity of the multilayered structure, the unambiguous determination of the $\chi(z)$ profile is not as straightforward as for chiral bulk magnets~\cite{Depth_profile_18,Binding_String_NL_22}, especially in combination with the variation of the incident angle.
This is due to the appearance of complex interference effects in the multilayers, combined with the necessity to not only include layer-dependent absorption, but also dispersion.
However, it is possible to obtain qualitative information about the $\chi(z)$ profile.
In general, at photon energies far away from the Fe $L_3$ absorption edge, the soft x-ray attenuation length increases, while it is shorter near the resonance.
Consequently, the top part of the multilayer which is closest to the surface contributes more to the total measured $\chi_\text{m}$.
In other words, if the $\chi(z)$-profile monotonically changes between $180^\circ$ and $0^\circ$ as a function of depth, the dividing vector will slightly rotate when changing the photon energy across the Fe $L_3$ edge. 
This behavior is consistent with our observations in Fig.~\ref{fig:CDREXS}(a). 
Furthermore, the sampling depth can be changed over a larger range by varying the incident x-ray angle $\theta$ ($\theta = 90^\circ - \alpha$, where $\alpha$ is the scattering angle defined with respect to the surface normal)~\cite{sm}.
As shown in Figs.~\ref{fig:CDREXS}(b) and \ref{fig:CDREXS}(c), by changing $\theta$ from $32^\circ$ to $35^\circ$, the CD-REXS pattern reverses sign, suggesting the existence of a smoothly-varying 3D hybrid domain wall structure.
Our CD-REXS results are therefore consistent with previous work on Ta/[CoFeB/MgO/Ta]$_{16}$~\cite{li2019anatomy}, and suggest \neel\ magnetic character near the surface and substrate, with Bloch-like character in the central region.
A plot of $\chi_\text{m}$ as a function of energy, extracted from the CD-REXS image stack in Fig.~\ref{fig:CDREXS}(a), is shown in Fig.~\ref{fig:CDREXS}(d).
Note that $\chi_\text{m}$ for larger energies is not defined, i.e., there is no clear dividing line in CD-REXS contrast as a result of the complex interference effects in the heterostructure.
It is therefore beneficial to employ additional techniques, such as micromagnetics and LE-\musr, to further understand changes in the magnetic states with depth.

\subsection{Micromagnetics}
\begin{figure}
	\centering
	\includegraphics[width=\linewidth]{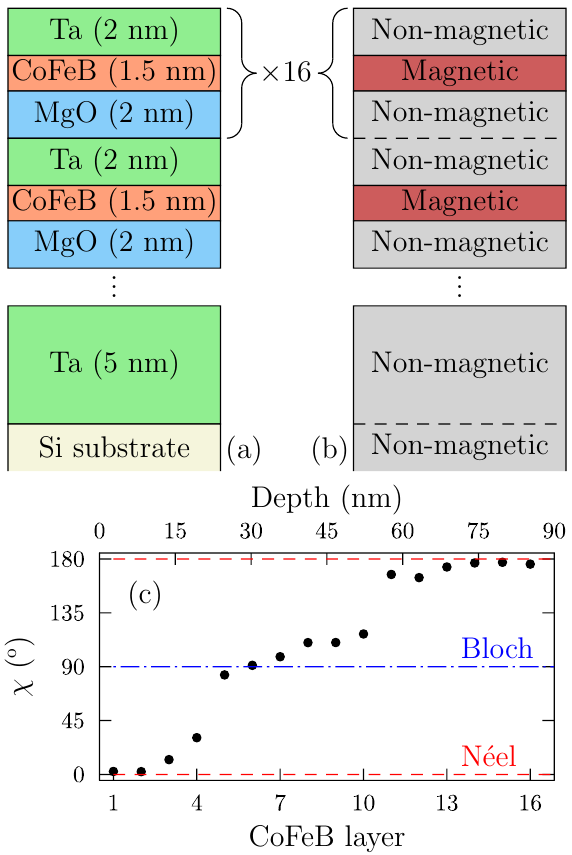}
	\caption{(a) A diagram of the multilayer sample used in this work, with a demonstration of how this has been converted to a micromagnetic model in (b). The helicity of the multilayer system, $\chi$, changes from \neel-type twisting at the top and bottom surfaces, to Bloch-like in the centre of the stack.}
	\label{fig:micromagnetics}
\end{figure}

We have performed micromagnetic simulations of our multilayer system.
Our sample has been converted to a micromagnetic model as shown in Fig.~\ref{fig:micromagnetics}(a--b); in all layers apart from CoFeB we set the saturation magnetisation to zero, meaning that the magnetic layers interact only through the long-range dipolar field between them.
We find that a magnetic state with chirality, in our case a skyrmion-like object, is lower in energy than a ferromagnetic configuration, suggesting that complex twisting magnetic configurations are the magnetic ground state.
By initialising the system with a skyrmion before allowing it to relax, we are able to straightforwardly evaluate $\chi$ as a function of depth; the results are shown in Fig.~\ref{fig:micromagnetics}(c).
We find agreement with Ref.~\cite{li2019anatomy}, specifically a \neel-type magnetic configuration on the top and bottom surfaces, with Bloch-like character in the middle of the stack.
We were unable to simulate a skyrmion system with depth-independent helicity, suggesting that the depth-dependence significantly reduces the energy of the system.
One might speculate that this depth-dependent helicity is inherent to a system that changes from semi-infinite (at the surfaces) to infinite (in the middle of the system) on the length scale of the interactions as it may minimise the free energy cost of the stray field.

\subsection{Muon-spin rotation}
With a consistent picture of the magnetism in Ta/[CoFeB/MgO/Ta]$_{16}$ arising from CD-REXS and micromagnetic simulations, we now turn to \musr, which is a powerful technique that allows one to learn about the magnetism from a local perspective.
However, it is often challenging to interpret results from \musr, especially in complex magnetic systems.
\musr\ depends intricately on the atomic configuration, with the location of the implanted muon with respect to the magnetic ions having significant impact on the field detected in a \musr\ experiment.
Conversely, micromagnetics, by design, ignores atomic details of a material in favor of a continuum approximation.
One would therefore not generally expect these two techniques to be complementary.
In this case however, due to the existence of large volumes of non-magnetic material in which the muon will also stop and predominately feel the long-range dipolar field, the continuum limit in these regions is likely to provide a good estimate of the experimentally observed magnetic fields.

In a muon-spin rotation (\musr) experiment~\cite{blundell1999spin,blundell2021muon} spin-polarized muons are implanted in a sample where they interact with the local magnetic field at the muon site.
After, on average, 2.2~$\mu$s, the muon decays into a positron and two neutrinos.
By detecting these positrons, which are preferentially emitted in the direction of the muon-spin at the time of decay, we can track the polarization of the muon-spin ensemble.

\begin{figure}
	\centering
	\includegraphics[width=\linewidth]{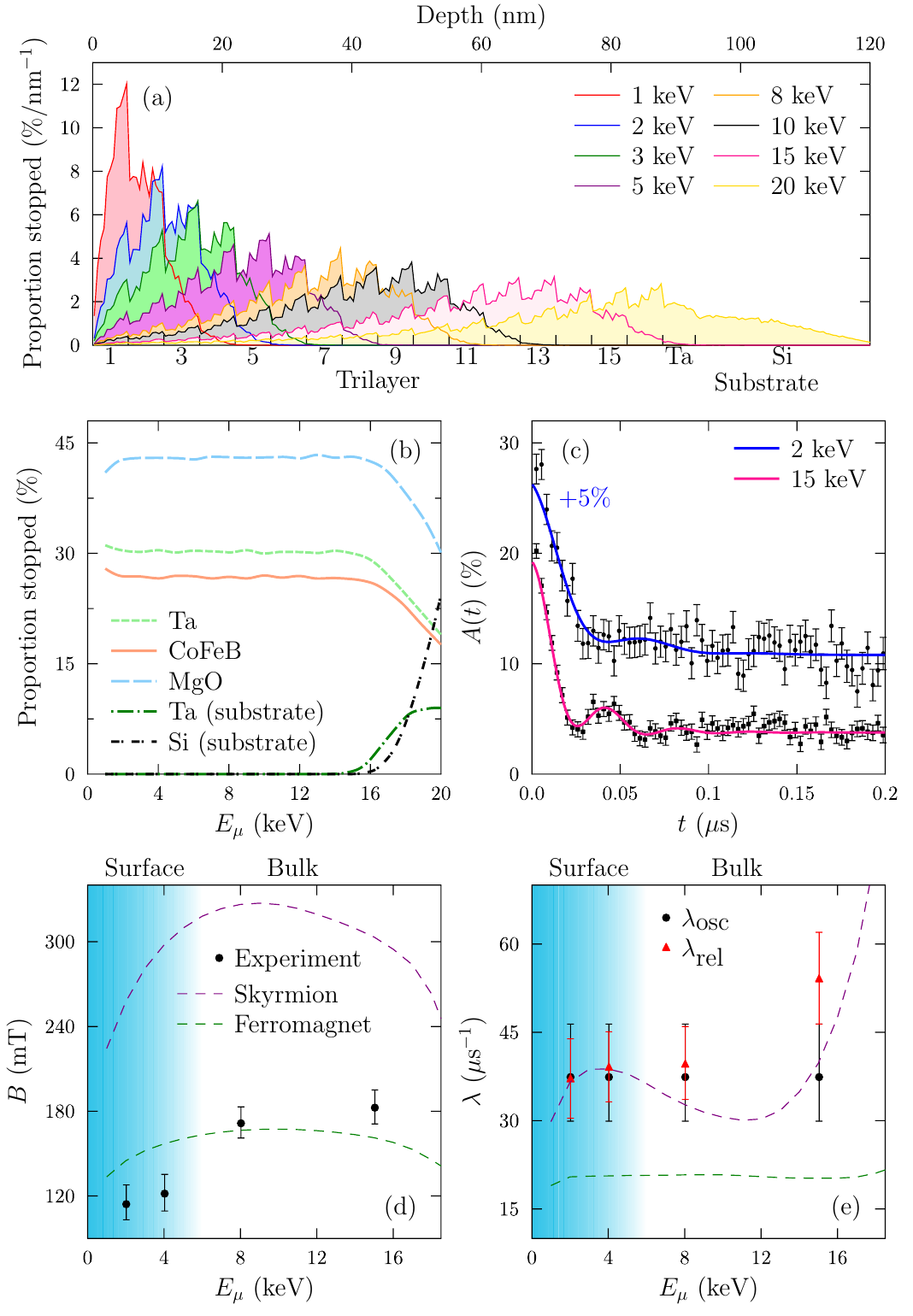}
	\caption{(a--b) Muon stopping profiles: (a) implantation energy dependence, (b) the proportion of muons stopping in each layer type. (c) Typical ZF \musr\ spectra $A(t)$ with parameters extracted from fitting shown as points in (d--e). The dashed lines in (d--e) show the results from micromagnetic simulations of a ferromagnetic and a skyrmion-like state.}
	\label{fig:muonZF}
\end{figure}

Simulations of the muon stopping profile in our sample performed using TRIM.SP~\cite{morenzoni2002implantation} [Fig.~\ref{fig:muonZF}(a)] demonstrate how the muons are distributed among the layers at different incident energies $E_\mu$.
Owing to the spread in the stopping profile, it is possible to distinguish the top layers from those deeper layers within the sample, although it is not possible to solely probe the layers of the sample nearest the substrate.
The proportion of muons stopping in each layer type is shown in Fig.~\ref{fig:muonZF}(b), and demonstrates that in the range $\simeq2$--$16$~keV the proportion of muons in magnetic/non-magnetic layers does not change.
In magnetic thin films, the dipolar field extends into nonmagnetic layers with a length scale determined by the size of magnetic modulations~\cite{tsymbal1994evaluation,krieger2019topology}; in our case, this is likely to be 10s of nm, matching the length scale of the skyrmions.
There are sufficient muons stopping in all layers to probe the magnetism throughout the sample.

In zero applied field (ZF) we expect the system to host relatively disordered helical magnetic order. 
ZF \musr\ measurements at $T=295$~K as a function of $E_\mu$ show the existence of two distinct regimes of behavior, classified by different implantation depths.
(For all measurements the muon spin is initially rotated approximately 10\degree\ out of the plane of the film.)
The differences with $E_\mu$ can be seen in the muon asymmetry $A(t)$, [Fig.~\ref{fig:muonZF}(c)].
To track the changes we parameterize the data~\cite{sm,pratt2000wimda,james1975minuit,suter2012musrfit} using 
\begin{align}
A(t)=&a_\text{osc}^\text{ZF}\exp\left(-\lambda_\text{osc} t\right)\cos\left(\gamma_\mu Bt+\phi\right)\nonumber\\
&+  a_\text{rel}^\text{ZF}\exp\left(-\lambda_\text{rel} t\right) + \sum_{i=3}^4a_{\text{bg},i}\exp\left(-\sigma_{\text{bg},i}^2t^2\right).
\end{align}
Many of these parameters are energy-independent~\cite{sm}, leaving the term with amplitude $a_\text{osc}^\text{ZF}$ containing the majority of the information about the variation of the magnetic properties with depth, and the terms with amplitudes $a_{\text{bg},i}$ predominantly capturing background effects.
The $a_\text{osc}^\text{ZF}$ term arises from muons that stop in sites where the muon spin and local field $B$ are not parallel, leading to coherent spin precession.
The relaxation rates depend both on the static distribution of fields at the muon site, and the dynamic fluctuations on the muon (MHz) timescale.
The change in $B$ with $E_\mu$, which can be unambiguously extracted from the data irrespective of parameterization, is shown in Fig.~\ref{fig:muonZF}(d); the internal field significantly increases from around 120~mT at 4~keV and below to around 180~mT at 8 keV and above.
Conversely, the relaxation rates [Fig.~\ref{fig:muonZF}(e)] remain approximately constant as a function of implantation energy.

There are two regimes evident in Fig.~\ref{fig:muonZF}(d), with the measurements at 4~keV and below showing different behavior to those at 8~keV and above.
Comparing with Fig.~\ref{fig:muonZF}(a), we suggest a surface regime is formed from approximately the first 4 stacks nearest the surface, corresponding to a penetration depth up to $\simeq22$~nm.
At higher $E_\mu$, the local field at the muon sites increases sharply in magnitude, suggesting a change in the magnetism.

To understand these changes, we compute the \musr\ spectra from our micromagnetic calculations (assuming no dynamic effects) for both a ferromagnetic and skyrmion-like configuration (that is, a skyrmion whose shape is affected by the finite size of the simulation) by weighting the calculated field at each depth by the relevant stopping probability for each $E_\mu$.
We are then able to extract the simulated internal field and relaxation rate by fitting the simulated spectra in the same way we fit the measured \musr\ spectra.
The magnetic configurations explored in our simulations are picked as the simplest ones that allow us to describe the magnetic states found experimentally, since they represent the limiting cases, and allow us to explore the effect of introducing chirality.
Specifically, the reported disordered helical groundstate of the material appears ferromagnetic on a short-length scale.
On a longer length scale it appears to host disordered and chiral regions more akin to the simulated skyrmion-like configuration.
The behaviour realized in the material is therefore anticipated to be something between the two simulations.
We note that, due to the limited system size possible in our micromagnetic simulations, it is not feasible to directly explore the disordered helical groundstate on the length scales on which it is found to exist in experiment.
However, we find that the results of small simulations of a helical state are consistent with those of our skyrmion-like configuration.

From our micromagnetic simulations we find that the calculated local field in the magnetic CoFeB layers is much higher and more varied than that in the non-magnetic layers.
This will lead to fast precession with a large relaxation rate. However, the inherent experimental time-resolution of our measurement means that it is unlikely that we could observe these effects.
Whilst the value of the field in the CoFeB layers in experiment is not expected to agree with the micromagnetic calculations (owing to this field being sensitively determined in the material by the local environment of the muon, which is not captured in a micromagnetic approach), we expect that a large, varied field will be experienced by the muon ensemble in these layers, hence preventing them from significantly contributing to the measured spectra.
We therefore suggest that the measured spectra predominantly probe the stray dipolar field in the non-magnetic layers that arises from the ordered magnetism in the magnetic CoFeB layers.

The magnetic field extracted from the ferromagnetic micromagnetic calculation is consistent with the magnitude of that extracted from the experimental ZF-\musr\ spectra.
The simulation shows a small increase in field as $E_\mu$ increases (likely to occur because the muon goes from seeing a ``half-infinite'' sample to an ``infinite'' one), although this increase is not as large as that observed experimentally.
Conversely, although the field is much higher in the skyrmion-like calculation, it does show an increase much closer to that observed experimentally.
The experimental relaxation rates, which are sensitive to the static distribution of magnetic fields seen by the muon, show little dependence on $E_\mu$.
Both simulated magnetic states predict relaxation rates with relatively little $E_\mu$-dependence.
However it is the relaxation from the skyrmion-like calculation whose order of magnitude is consistent with experiment in this case.
(We also note that the disorder in the ferromagnetic calculation, and hence $\lambda$, is likely to be larger than in a well-ordered ferromagnetic sample due to the edge effects contributing significantly at this simulation size.)
Taken together, these results imply that the material hosts a magnetic state that has some character of both calculations, not unlike the somewhat disordered helical state we expect.
The large measured relaxation rate suggests a wide distribution of local magnetic fields at the muon site, consistent with a complex, twisting magnetic structure.
Further, the significant increase in $B$ with $E_\mu$ can only be explained by the micromagnetic simulation of the skyrmion-like configuration, where $\chi$ is depth-dependent.
Moreover, our simulations demonstrate that the precise value of the \musr-fitting parameters will depend sensitively on the proportion of the sample that is domain wall- or ferromagnetic-like.
Our \musr\ results are therefore consistent with micromagnetic simulations and CD-REXS, and suggest a quantitative agreement between the expected depth of changes in the magnetism.

\begin{figure}
	\centering
	\includegraphics[width=\linewidth]{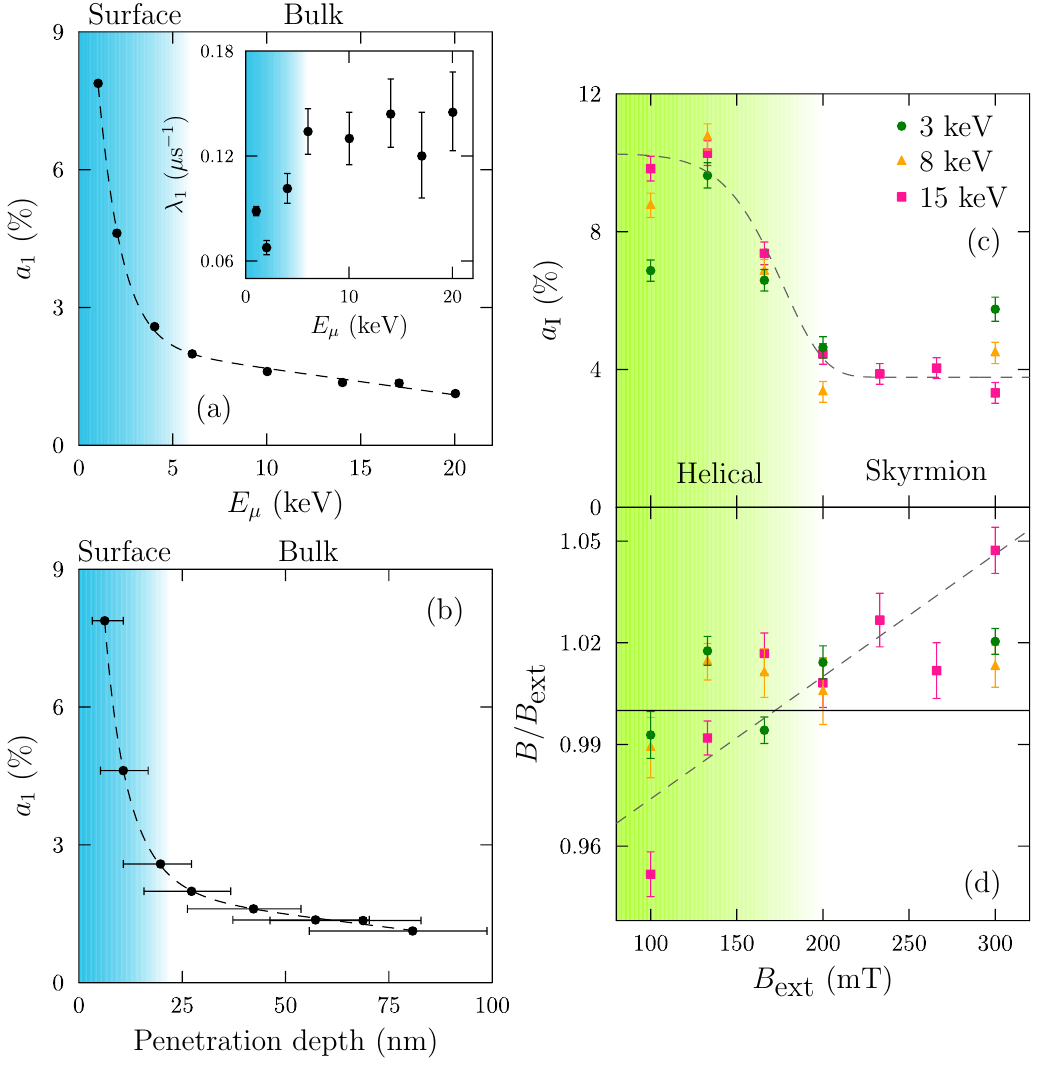}
	\caption{(a) Energy dependence of parameters from fitting TF \musr\ at 10~mT, with a conversion to average muon penetration depth in (b). Error bars in (b) represent the asymmetric range in which 2/3rds of the muons stop according to the calculations in Fig.~\ref{fig:muonZF}(a). (c--d) Field-dependent parameters from fitting TF \musr\ data, showing (c) the amplitude $a_\text{I}$, and (d) the  ratio of $B$ to the applied field $B_\text{ext}$. Dashed lines are guides to the eye.}
	\label{fig:muonTF}
\end{figure}

To explore these changes further, TF \musr\ measurements were performed at 295~K as a function of $E_\mu$ in an applied magnetic field of 10~mT, applied perpendicular to the film, for which we expect the magnetic state to be relatively undisturbed.
Representative TF \musr\ data can be found in the Supplemental Material~\cite{sm}.
For these data we fit the number of counts in each detector simultaneously, using the polarization function
\begin{equation}
P_x(t)=\sum_{i=1}^2a_i\exp(-\lambda_it)\cos(\gamma_\mu Bt+\phi),
\end{equation}
where, due to the difficulty in resolving two different fields at the muon site (in particular due to the large $\lambda_2$), and as we expect the shift in measured field to be small in magnetic thin films~\cite{salman2012proximal}, we constrain the fits to share $B$ between the two terms.
We find many parameters to be energy-independent~\cite{sm}.
The energy-dependent parameters, $a_1$ and $\lambda_1$, are shown in Fig.~\ref{fig:muonTF}(a).

We interpret the two different terms as arising from muons stopping at different classes of muon site in the sample.
Since there is little change in the proportion of muons stopping in each different material as a function of $E_\mu$ [Fig.~\ref{fig:muonZF}(b)], the results suggest the magnetic structure must change as a function of depth.
The site contributing to the term with amplitude $a_1$ experiences a small internal field and has a small relaxation rate, leading to coherent precession in the applied magnetic field.
This suggests these sites are sensitive mainly to regions of disordered magnetic field (where the average magnetic field cancels), such as domain wall-like regions.
This weak relaxation rate $\lambda_1$ [inset of Fig.~\ref{fig:muonTF}~(a)] increases with depth.
Whilst it is tempting to interpret this in terms of the changes in the static magnetism that are supported by the results of our micromagnetic simulations (discussed above), it might be more reasonable to attribute such a weak relaxation rate [which is two orders of magnitude smaller than measured and simulated in Fig.~\ref{fig:muonZF}(e)] to dynamics in the system.
In this case, the increase in $\lambda_1$ would, in a fast-fluctuation limit, suggest that the frequency of dynamic fluctuations decreases at these sites with depth.
Such behaviour could be expected since surface-dominated states have dynamics arising from rapid fluctuations on a relatively-short lengthscale (as the system tries to find a configuration that satisfies the constraint of the surface) to one where longer wavelength collective excitations dominate in the bulk of the material.
As $E_\mu$ increases, the amplitude $a_{1}$ decreases, consistent with fewer surface-state muon-sites being realized with depth.
We again see a crossover around trilayer 4 of the material.
Although this appears relatively smooth in Fig.~\ref{fig:muonZF}(a), this at least partially reflects the fact that the stopping profile is spread across several trilayers at each implantation energy.
To elucidate this further, we have converted the implantation energy to a penetration depth using the results shown in Fig.~\ref{fig:muonZF}(a); and present $a_1$ again as a function of this parameter [Fig.~\ref{fig:muonTF}(b)].
We observe that $a_1$ remains approximately constant once the majority of muons are penetrating beyond around trilayer 4.
The changes observed in the TF measurements coincide with the change in magnetic structure seen from the ZF results, implying that the same crossover with depth is responsible for both measurements.
We therefore have a picture of (i) a static magnetic structure changing with depth (from those muon sites in large local magnetic field giving rise to the ZF signal), and (ii) the number of domain wall sites decreasing with depth, with the possibility that the dynamic fluctuation rate on the muon-timescale in these sites also decreases (from muon sites in regions with small local field giving rise to the wTF signal).
For both types of muon site, the environment changes rapidly around trilayer 4.

To access the region of the phase diagram where skyrmions are stabilized as the majority phase, further TF \musr\ measurements were performed as a function of larger magnetic fields, at $T=295$~K using implantation energies: 3, 8, and 15~keV.
Note that the thickness of both the magnetic and non-magnetic layers strongly affects the field at which skyrmions are stabilised (see Ref.~\cite{santos2023micromagnetic}), which is a result we are also able to reproduce with our micromagnetic simulations.
As such, the exact crossover field should be expected to be sample dependent.
For these data $P_x(t)=a_\text{I}\exp(-\lambda_\text{I}t)\cos(\gamma_\mu Bt+\phi_\text{B})+a_\text{II}\exp\left(-\sigma^2_\text{II}t^2\right)$ is found to describe the behavior over the entire dataset.
Most parameters are found to be independent of applied magnetic field or arising from background effects~\cite{sm}, hence we focus on the term with amplitude $a_\text{I}$, arising from muons stopping in sites where the spin precess in a local field $B$, which is found to be close to the applied field.
This implies we are again sensitive to the domain-wall sites where the average local field is small in these measurements.
The field-dependent changes in $a_\text{I}$ and $B$ are shown in Fig.~\ref{fig:muonTF}(c--d).

The proportion of muons stopping in the sample and giving coherent spin precession decreases as $B_\textrm{ext}$ is increased [Fig.~\ref{fig:muonTF}(c)].
This effect is approximately twice the size of that observed as a function of implantation energy in the low-TF measurements, suggesting that applied magnetic field leads to a more rapid decrease in the occurrence of these domain-wall environments compared to their variation with depth.
This picture is supported by the results of x-ray ptychography measurements~\cite{li2019anatomy} that suggest just such a change in magnetic structure with out-of-plane magnetic field, such that skyrmions become the dominant magnetic defect above $\sim$150~mT.
The skyrmions represent a limiting case of the domain-wall structure, minimising the sample volume occupied by the low-field muon stopping sites.
This is consistent with our measurement in which, around 180~mT, the amplitude parameter ceases changing, suggesting a crossover to a different regime.
We therefore suggest that this crossover can be seen at all implantation energies, and that the plateau of $a_\text{I}$ indicates the stabilization of skyrmions at each depth within the stack that we have probed.

There appears to be a weak dependence of the ratio $B/B_\text{ext}$ with $B_\text{ext}$, Fig.~\ref{fig:muonTF}(d).
As $B_\text{ext}$ increases, the enhancement of $B$ becomes larger, from predominantly below the applied field, to  above, most pronounced at 15~keV.
The crossover between these regions appears to approximately coincide with the transition to skyrmion-dominated magnetic structures.
The micromagnetic simulations suggest that the skyrmion-like states will exhibit a higher field at the muon site than when the system is ferromagnetic; it is therefore expect that crossover from a disordered helical state (which appears almost ferromagnetic on a short-length scale) to a disordered skyrmion state might also show a similar, albeit weaker, effect.
Further, the reduction in $a_\text{I}$ in the skyrmion-stabilized phase possibly reflects the behaviour of muons in sites where the spin is relaxed too rapidly to be detected, reducing $a_\text{I}$.
This is consistent with the response seen in bulk skyrmion systems~\cite{franke2018magnetic,hicken2020magnetism,hicken2021megahertz}, where the relaxation rate is enhanced at the center of the skyrmion lattice phase, even when the skyrmion lattice is not the majority phase.

\section{Conclusion}
In summary, our LE-\musr\ and CD-REXS measurements identify changes in the magnetism of the $[$Ta(2)/CoFeB(1.5)/MgO(2)$]_{16}$/Ta(5) multilayer stack on a Si substrate as a function of both depth and applied field.
Our micromagnetic simulations well explain these observations, and we have demonstrated the utility of cheap micromagnetic simulations for explaining \musr\ results when there is a large volume fraction of the sample that is non-magnetic.
CD-REXS demonstrates the continuous evolution of the helicity angle with depth, while \musr\ shows that the magnetism in the top trilayers of the multilayer stack have a different static magnetic structure to those further from the surface.
The trilayer at which this crossover is observed provides quantitative agreement with predictions from simulations~\cite{li2019anatomy,CoTb_PRL_21}.
We also identify a decrease in the volume occupied by domain walls with both depth and applied magnetic field, with the latter undergoing a crossover to a regime dominated by skyrmions around 180~mT.
Our work highlights the utility of the LE-\musr\ technique in studying multilayer skyrmion systems, providing unique insights into the depth dependence of the magnetic textures and their properties.
Out findings confirm results from computational predictions, and paves the way for future investigations to experimentally verify a number of other computational works on technological applications of skyrmion-hosting systems.
Determining the depth-dependence of the magnetic properties of multilayered systems is of paramount importance for their incorporation in device applications.

\section*{Acknowledgments}
Part of this work was carried out at the Paul Scherrer Institut, Switzerland; we are grateful for the provision of beamtime.
The authors acknowledge Diamond Light Source for beamtime on beamline I10 under proposal SI-18898.
The project was funded by EPSRC (UK) (Grant No: EP/N032128/1).
M.N. Wilson acknowledges the support of the Natural Sciences and Engineering Research Council of Canada (NSERC).
Research data from this paper will be made available via Durham Collections at https://doi.org/10.15128/r2qn59q403m.

\bibliography{bib}

\end{document}